# Topological Rainbow Trapping for Spatial-frequency Demultiplexing of Underwater Acoustic Signals

**Cheng Lin [a], Yangkai Liu [a], Tuo Liu [b], Yi Zhang [c], Haiyan Fan [a, *], Hui Zhang [a, *]**

[a] *Jiangsu Key Laboratory for Design and Manufacturing of Precision Medicine Equipment, School of Mechanical Engineering, Southeast University, Nanjing, Jiangsu 211189, China*

[b] *State Key Laboratory of Acoustics and Marine Information Institute of Acoustics, Chinese Academy of Sciences, Beijing, 100190, China*

[c] *Shenzhen Research Institute of Big Data, The Chinese University of Hong Kong (Shenzhen), Shenzhen, Guangdong 518000, China*

## Abstract:

Efficient separation and localization of multifrequency acoustic waves are essential for underwater target recognition and acoustic energy harvesting. Topological rainbow trapping enables frequency demultiplexing by combining slow-wave localization with topologically protected transport, yet its underwater implementation remains challenging because of complex fluid–solid interactions and the difficulty of integrating long-range transport with frequency-selective localization in an open system. Here, we theoretically develop and experimentally demonstrate two underwater spatial–frequency demultiplexing mechanisms based on the acoustic analogues of the quantum valley Hall effect (QVHE) and quantum spin Hall effect (QSHE). Both employ spoof surface acoustic waves (SSAWs), whose fields are confined near a structured surface and decay evanescently into the surrounding water, enabling experiments without an enclosed waveguide. In the QVHE mechanism, a spatial gradient along a valley-Hall edge channel shifts the local edge-state dispersion, causing different frequencies to accumulate at distinct positions and thereby realizing spectral and spatial demultiplexing. In the QSHE mechanism, one-dimensional topological edge states are coupled to frequency-selective zero-dimensional higher-order corner states. Multifrequency signals first propagate robustly along a common boundary and are then transferred to prescribed remote corners according to frequency, producing a transport-then-confinement process. This mechanism combines defect-tolerant edge transport, frequency-selective corner localization, and remote rainbow trapping. Numerical simulations and measurements verify the frequency-dependent localization and the persistence of the designed transport pathways under structural defects. The proposed open SSAW platform performs robust frequency demultiplexing at the physical layer, reducing reliance on digital signal processing and offering potential for underwater target recognition and frequency-selective acoustic energy harvesting.

*Corresponding author.

*E-mail address:* Chenglin@seu.edu.cn (C. Lin), yangkai.liu@seu.edu.cn (Y. Liu), liutuo@mail.ioa.ac.cn (T. Liu), zhangyi@sribd.cn (Y. Zhang), haiyan.fan@seu.edu.cn (H. Fan), seuzhanghui@seu.edu.cn (H. Zhang)

# 1. Introduction

In complex and dynamic underwater environments, the severe attenuation of electromagnetic waves renders acoustic waves the primary medium for long-range underwater communication and target detection [1–5]. The radiated noise of surface vessels, submarines, and increasingly prevalent unmanned underwater vehicles generally contains multiple frequency components [6]. Isolating diagnostically relevant frequencies from such mixed signals is therefore essential for reliable underwater target recognition. Moreover, the marine environment is rich in acoustic wave energy. Harvesting this energy through piezoelectric [7], resonant [8], or nanostructured designs [9,10] could support the development of self-powered underwater sensor networks. Underwater acoustic metamaterials [11,12], which have advanced rapidly over the past decade, provide flexible means of manipulating waterborne acoustic waves and offer new opportunities for both target recognition and energy harvesting. Through artificially engineered subwavelength structures, these metamaterials can exhibit effective acoustic properties that are difficult to obtain in natural media, enabling precise control of underwater acoustic wave propagation at the subwavelength scale. In recent years, underwater acoustic metamaterials have demonstrated tremendous potential in fields such as sound absorption [13–16], acoustic wave focusing [17–19], acoustic cloaking [20–22], and topological acoustics [23–25].

Among these developments, topological acoustic systems are particularly attractive because they can support boundary modes with suppressed backscattering [26,27], as well as higher-order localized modes [28] associated with nontrivial topological invariants [29–32]. More recently, spatially graded topological structures have been introduced to establish a position-dependent relationship between frequency and wave localization, giving rise to the topological rainbow effect [33]. In a spatially graded topological structure, the local dispersion evolves along the propagation direction, causing different spectral components to approach their respective slow-wave or stopping conditions at distinct spatial locations [34]. This process enables spectral demultiplexing, spatial separation, and selective wave localization [33,35,36]. Accordingly, the topological rainbow effect has been demonstrated in various physical fields, such as photonic systems [35–38], airborne acoustic systems [39,40], and elastic wave systems [41].

Transferring this concept to waterborne acoustics is nontrivial. The coupled dynamics of the liquid medium and solid structure modify modal dispersion and field confinement, increasing the complexity of both structural design and experimental implementation. Consequently, underwater realizations of topological rainbow trapping remain comparatively limited. Furthermore, beyond realizing rainbow localization underwater, it remains difficult to construct an open underwater acoustic system that combines topologically protected transport over extended propagation distances with spectral discrimination and subsequent energy confinement at prescribed remote sites. Addressing these challenges could shift preliminary spectral processing from the digital domain to the acoustic hardware itself, thereby reducing dependence on active control systems and downstream signal-processing algorithms. Such functionality would be valuable for the extraction of characteristic frequency components, underwater target recognition, and spatially distributed acoustic energy harvesting.

In this paper, we theoretically develop and experimentally demonstrate topological rainbow trapping in an open underwater environment based on spoof surface acoustic waves (SSAWs), as illustrated in Fig. 1. The surface-bound nature of SSAWs permits acoustic manipulation in an unenclosed water domain, thereby providing a suitable system for implementing topological rainbow trapping without an enclosing acoustic cavity. The system consists of two spatially graded two-dimensional (2D) topological acoustic configurations designed to support valley-Hall edge-state trapping and pseudospin-mediated edge-to-corner localization, respectively. The upper configuration is based on an acoustic analogue of the quantum valley Hall effect (QVHE) and exploits valley-polarized topological edge states along a spatially graded interface. The gradient continuously modifies the local edge-state dispersion, causing different frequency components to reach their respective stopping conditions at distinct positions along the propagation path. This frequency-dependent position mapping enables joint spectral and spatial demultiplexing. The lower configuration is based on an acoustic analogue of the quantum spin Hall effect (QSHE), in which pseudospin-polarized one-dimensional (1D) edge states are coupled to frequency-selective zero-dimensional (0D) higher-order corner states. Acoustic signals are first transported over an extended distance along the topological boundary and are subsequently coupled into remote corner modes whose resonance frequencies match the corresponding spectral components. This edge-to-corner coupling produces a “transport-then-confinement” process, enabling frequency-selective remote localization following long-range topological edge transport. Numerical simulations incorporating structural defects further indicate that the edge-guided waves bypass the perturbations and continue to excite the prescribed remote corner states, demonstrating the defect tolerance of the edge-to-corner transport pathway under the tested conditions. Theoretical analyses and experimental measurements demonstrate position-dependent spectral separation in the QVHE configuration and defect-tolerant edge-to-corner rainbow trapping in the QSHE configuration. By combining protected boundary transport with geometrically prescribed corner localization, the open SSAW system provides a physical-layer strategy for underwater frequency demultiplexing and offers potential for target recognition and broadband acoustic energy harvesting.

## 2. Model and Methods

## 2.1 Unit-cell dispersion relations and topological phase transition

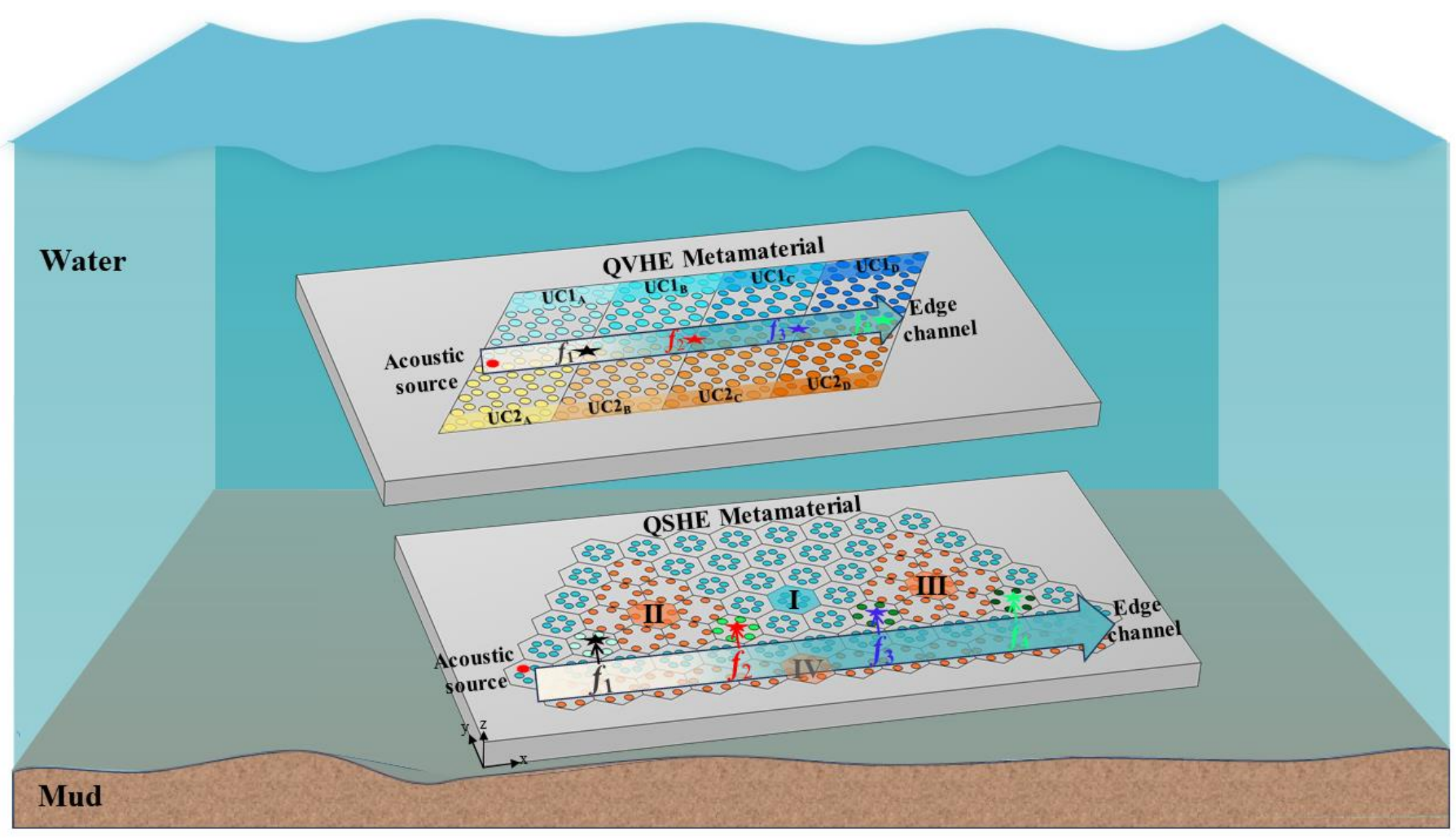

**Fig.1.** The open underwater SSAW system for spatial-frequency demultiplexing and edge-to-corner topological rainbow trapping.

Figure 2(a) illustrates the geometric configuration of the QVHE unit cell, with a lattice constant of $a$ = $0.7\sqrt{3}$ mm. The structure consists of an aluminum plate with a thickness of $l$ = 0.9 mm immersed in water, in which two circular blind holes of depth of $l_0$ = 0.6 mm are arranged in a honeycomb pattern. The centers of the two holes are positioned symmetrically along the long diagonal of the unit cell. The reference size parameter is defined as $d$ = 0.4 mm. To characterize the topological phase transition induced by the broken spatial inversion symmetry, a dimensionless perturbation parameter $\delta$ is introduced to tune the two holes according to $d_1 = d(1.3+\delta)$ and $d_2 = d(1.3-\delta)$, respectively. At $\delta$ = 0, the two sublattice holes are identical and the unit cell possesses $C_{6v}$ symmetry. Varying $\delta$ [for example, $\delta$ = ±0.1 in Fig. 2(a)] provides control over the bandgap. The dispersion relations are obtained from the eigenfrequency problem of the fully coupled solid–fluid system. For the solid (fluid) medium, the acoustic impedance is $Z_{AL} = \rho_{AL}c_{AL} = 16.821 \times 10^6\ \mathrm{kg\,m^{-2}\,s^{-1}}$ ( $Z_W = \rho_W c_W = 1.49 \times 10^6\ \mathrm{kg\,m^{-2}\,s^{-1}}$ ). The resulting impedance ratio of approximately 11.3 is not sufficiently large to justify a perfectly rigid boundary approximation. The elastic response of the plate and the pressure field in water are therefore solved simultaneously, with continuity of normal motion and traction imposed at each solid–fluid interface.

Bloch–Floquet periodic conditions are applied to the in-plane boundaries, and the Bloch wave vector is swept along the Γ–K–M–Γ path of the first Brillouin zone. In Fig. 2(b), the gray shaded regions represent the projected continuum of propagating acoustic modes in water, delimited by the water sound lines. Modes within these regions can radiate energy into the surrounding water and therefore do not constitute tightly confined SSAW modes. For the inversion-symmetric cell with $\delta = 0$, the two bands shown by the black dotted curves meet at a twofold-degenerate Dirac point at the K valley, as shown in Fig. 2(b). When $\delta > 0$ or $\delta < 0$, the geometric deformation breaks the spatial inversion symmetry of the lattice leading to the lifting of the Dirac degeneracy and formation of a bandgap, as depicted by the red and blue curves in Fig. 2(b).

Figure 2(c) shows the evolution of the two K-point eigenfrequencies as $\delta$ varies continuously from −0.1 to +0.1. The bandgap narrows as $\delta$ approaches zero, closes at the Dirac point when $\delta = 0$, and reopens after the perturbation changes sign. Although the unit cells with $\delta = 0.1$ and $\delta = -0.1$ exhibit identical dispersion relations under opposite perturbation signs, their underlying topological properties are distinct. As evidenced by Fig. 2(c), the dominant pressure response is exchanged between the two inequivalent blind holes at the K point when the sign of δ is reversed, providing direct evidence of band inversion.

The band inversion is accompanied by a reversal of the valley topology. This signifies that the propagating modes within the two structures possess distinct topological valley indices. The topological distinction between the two phases can be characterized by the valley Chern index. Owing to time-reversal symmetry, the Berry curvatures near the inequivalent K and K′ valleys have opposite signs, so the total Chern number over the entire Brillouin zone remains zero [42]. Integration of the Berry curvature around a single valley for the band below the gap gives $C_K = -1/2$ for $\delta < 0$, indicating a topologically trivial phase, $C_K = +1/2$ for $\delta > 0$, characterizing a topologically nontrivial phase (see Note 2 in Supplementary Materials).

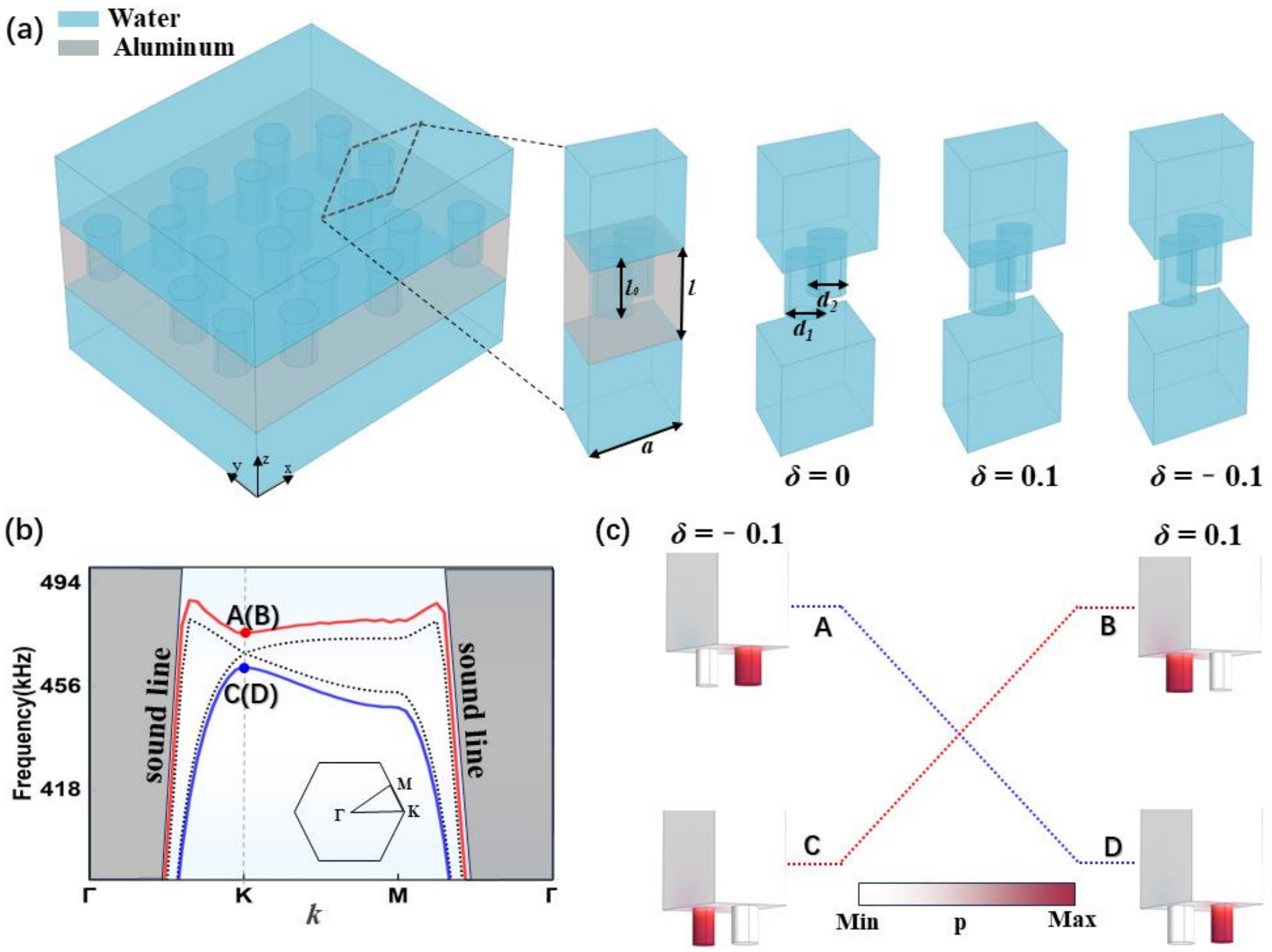

**Fig. 2.** Schematic of the underwater QVHE unit cell and corresponding dispersion relations. (a) Three-dimensional geometry of the aluminum–water unit cell and representative configurations for $\delta = 0$ and $\delta = \pm 0.1$. Gray and blue denote aluminum and water, respectively; $d_1$ and $d_2$ are the diameters of the two blind holes. (b) Dispersion relations along Γ–K–M–Γ. The black dotted curves correspond to the inversion-symmetric cell with $\delta = 0$, whereas the red

and blue solid curves show the upper and lower bands for the perturbed cells with $\delta = \pm 0.1$. The gray shaded regions represent the water-radiation continuum associated with the sound lines; the inset shows the first Brillouin zone and the selected wave-vector path. (c) Evolution of the K-point eigenfrequencies with $\delta$ and acoustic pressure distributions, demonstrating the exchange of modal character and the associated band inversion

To enable 0D acoustic localization, a composite QSHE unit cell consists of three primitive QVHE unit cells is therefore introduced, as shown in Fig. 3(a). The QSHE unit cells use through-holes instead of blind holes. In the unperturbed geometry, the six holes are arranged with sixfold rotational symmetry. Due to the band folding mechanism[43], the single Dirac point originally located at the $\mathbf{K}$ and $\mathbf{K}'$ valleys is mapped to the $\mathbf{\Gamma}$ point, leading to a fourfold-degenerate double Dirac point, as indicated by the black dashed line in Fig. 3(b). A radial displacement parameter $\Delta d$ is introduced to tune the positions of the six holes relative to the cell center: $\Delta d > 0$ corresponds to outward expansion, whereas $\Delta d < 0$ corresponds to inward contraction

The dispersion relations are calculated using the same fully coupled solid–fluid eigenfrequency framework as that used for the QVHE unit cell. As shown in Fig. 3(b), both the expanded configuration with $\Delta d = 0.15$d and the contracted configuration with $\Delta d = -0.26$d lift the fourfold degeneracy and open a bandgap around the double-Dirac frequency. The red and blue solid curves represent the bands above and below the opened gap for the perturbed configurations, while the labels A(B) and C(D) identify the corresponding nearly degenerate mode pairs at Γ. The evolution of the Γ-point eigenfrequencies and the associated pressure eigenfields are shown in Fig. 3(c). As $\Delta d$ is varied from negative to positive values, the bandgap closes at the double Dirac point and subsequently reopens with an inverted modal ordering. For the expanded cell with $\Delta d = 0.15d$, the upper mode pair A exhibits dipole-like $p_x$ and $p_y$ characteristics, whereas the lower pair C exhibits quadrupole-like $d_{x^2-y^2}$ and $d_{xy}$ characteristics. For the contracted cell with $\Delta d = -0.26d$, the order is reversed: the quadrupole-like pair B lies above the gap and the dipole-like pair D lies below it. This exchange of the *p*- and *d*-type mode characters provides direct evidence of band inversion and indicates a change in the sign of the effective Dirac mass. The *p*–*d* band inversion therefore marks a transition between two acoustic pseudospin-Hall phases. The spin-Chern analysis gives a vanishing spin Chern number for the contracted configuration ($\Delta d < 0$), identifying it as topologically trivial, and a nonzero spin Chern number for the expanded configuration ($\Delta d > 0$), identifying it as topologically nontrivial (see Note 3 in Supplementary Materials).

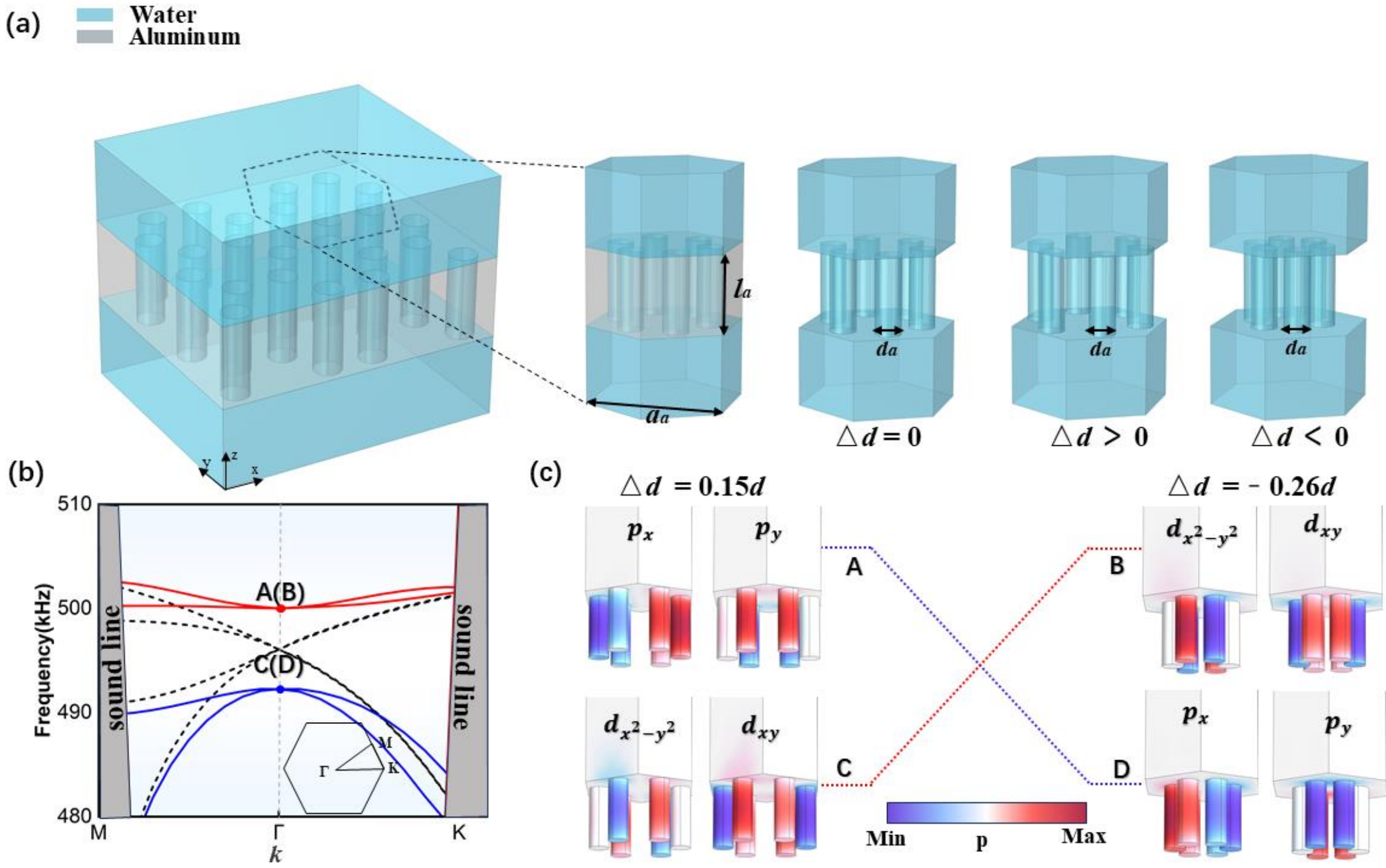

**Fig. 3.** Schematic of the composite QSHE unit cell and dispersion relations. (a) Three-dimensional geometry of the enlarged unit cell formed by three primitive QVHE cells, together with the unperturbed ($\Delta d = 0$), expanded ($\Delta d > 0$), and contracted ($\Delta d < 0$) configurations. Gray and blue denote aluminum and water, respectively. (b) Dispersion relations along M–Γ–K. The black dashed curves show the unperturbed double Dirac point at Γ, whereas the red and blue solid curves show the bands above and below the gap opened by the geometric perturbation. The gray shaded regions indicate the projected continuum of propagating waterborne modes; the inset shows the reduced Brillouin zone and the selected wave-vector path. (c) Evolution of the Γ-point eigenfrequencies and acoustic pressure distributions of modes A–D for $\Delta d = 0.15d$ and $\Delta d = -0.26d$, demonstrating the inversion between the dipole-like $p$ modes and quadrupole-like $d$ modes.

## 2.2 Supercell interface states and boundary bandgaps

Owing to their opposite valley Chern indices, the two QVHE domains support one-dimensional interface states within their common bulk bandgap. These states enable robust wave transport along the interface with suppressed backscattering, provided that intervalley scattering remains weak. Figure 4(a) shows a ribbon supercell formed by adjoining the QVHE unit cells UC1 ($\delta = -0.1$) and UC2 ($\delta = +0.1$), which possess opposite valley Chern indices. The periodic arrangement in the transverse direction produces two zigzag interfaces, denoted ZZ1 and ZZ2. In the dispersion relations in Fig. 4(b), the gray regions represent the bulk bands, whereas the red and blue branches correspond to edge states localized at ZZ1 and ZZ2, respectively. Both branches lie in the common bulk bandgap and connect the bulk bands, confirming the existence of valley-dependent interface channels. Figure 4(c) shows the acoustic pressure and elastic-displacement fields of modes E1 and E2 at the wave vectors marked in Fig. 4(b). Both modes are localized near the corresponding zigzag interfaces and decay into the adjacent domains. The co-localization of the scalar pressure field in water and the vector displacement field in aluminum demonstrates the simultaneous confinement of the coupled acoustic–elastic response by the QVHE interface states. Their distinct field symmetries are consistent with the reversed domain ordering at ZZ1 and ZZ2.

To establish the relationship between frequency and spatial position, we investigate the influence of the size tuning parameter $\delta$ on the dispersion characteristics. Figure 4(d) plots the dispersion relations of ZZ1 interface states across various $\delta$ (see Note 4 in supplementary materials for complete dispersion relations). It can be observed that the boundary state stably exists within the bandgap region, regardless of how $\delta$ changes. Furthermore, increasing $\delta$ systematically shifts the eigenfrequencies to higher regimes. This demonstrates that $\delta$ effectively modulates the frequency of the interface states.

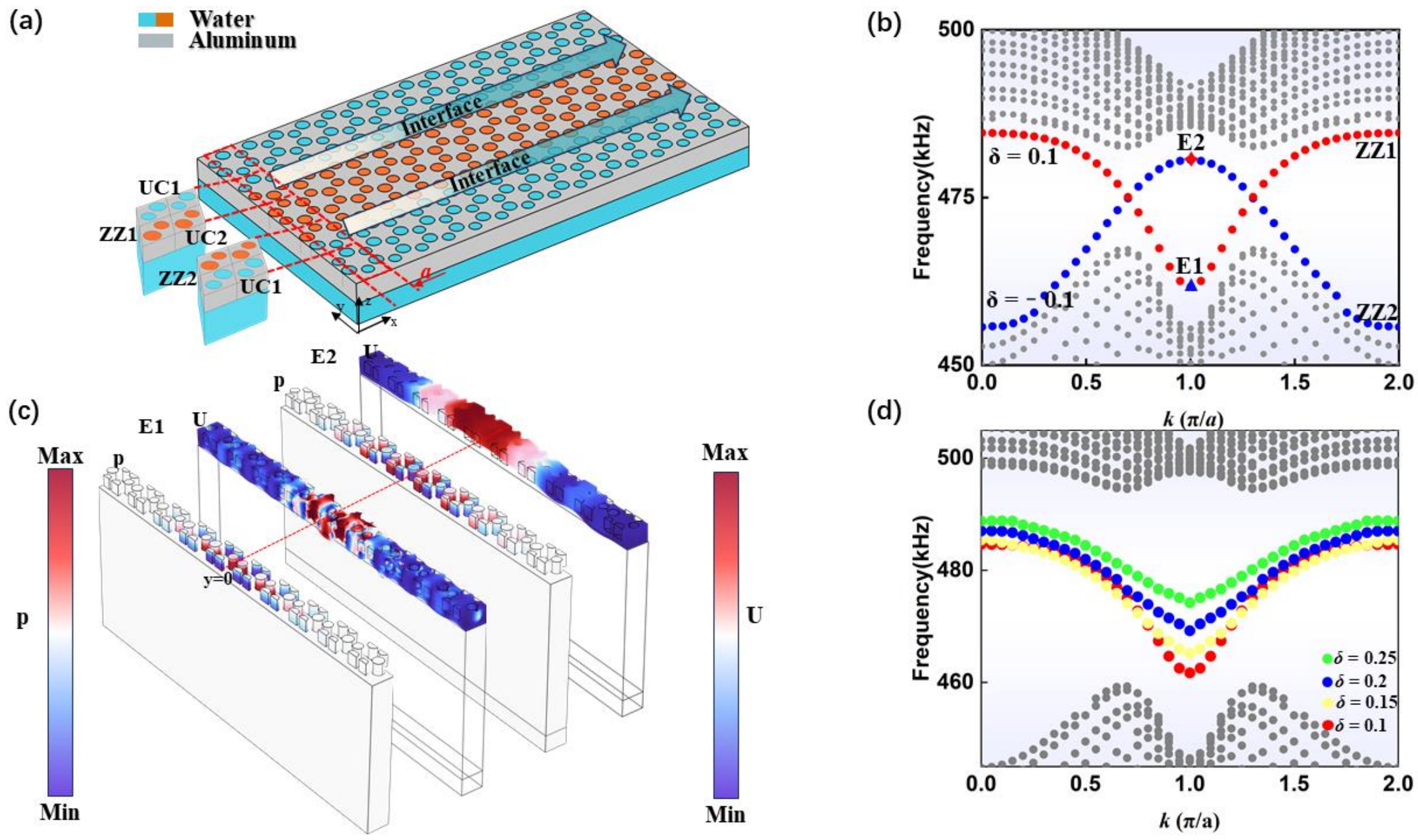

**Fig. 4.** QVHE ribbon supercell and dispersion relations. (a) Ribbon supercell composed of UC1 ($\delta = -0.1$) and UC2 ($\delta = 0.1$), forming two zigzag interfaces, ZZ1 and ZZ2. (b) Dispersion relations of the supercell. The gray regions denote the bulk bands, while the red and blue branches represent the interface states localized at ZZ1 and ZZ2, respectively. E1 and E2 indicate the representative eigenmodes examined in (c). (c) Scalar acoustic pressure fields in water and vector elastic-displacement fields in the aluminum plate for modes E1 and E2. Their co-localization near the corresponding interfaces demonstrates the simultaneous confinement of the coupled acoustic and elastic responses. (d) Evolution of the ZZ1 interface-state dispersion for $\delta = 0.1$, 0.15, 0.2, and 0.25, showing an upward frequency shift with increasing $\delta$.

Although the aforementioned valley-dependent supercell supports 1D topological interface states and enables frequency tuning, the acoustic energy remains confined to a 1D path. To further achieve 0D localization of acoustic energy, we analyze a composite QSHE supercell composed of the QSHE unit cell. a ribbon supercell is constructed from the contracted unit cell CUC1 ($\Delta d = -0.26d$) and the expanded unit cell CUC2 ($\Delta d = +0.15d$), as shown in Fig. 5(a). These unit cells correspond to topologically trivial and nontrivial pseudospin-Hall phases, respectively. The transverse periodicity of the ribbon produces two armchair interfaces with opposite domain orderings, labeled AM1 and AM2.

Figures 5(b) and 5(c) show the dispersion relations of the supercells containing the AM1 and AM2 interfaces, respectively. The gray circles represent bulk states, while the red and blue circles represent the topological interface state. Unlike the gap-connecting QVHE branches in Fig. 4(b), each QSHE edge branch spans only part of the common bulk bandgap. The AM1 branch occupies the lower part of the common bulk bandgap, leaving an upper edge bandgap. Conversely, the AM2 branch occupies the upper part leaving a lower edge bandgap. Consequently, edge bandgaps are formed [blue shadowed regions in Figs. 5(b) and 5(c)], within which localized 0D corner states can emerge. Figure 5(d) shows the underwater acoustic pressure distributions and displacement fields of the eigenmodes marked as CE1 in Fig. 5(b) and CE2 in Fig. 5(c). Their co-localization confirms the simultaneous confinement of the scalar acoustic and vector elastic responses by the QSHE edge states.

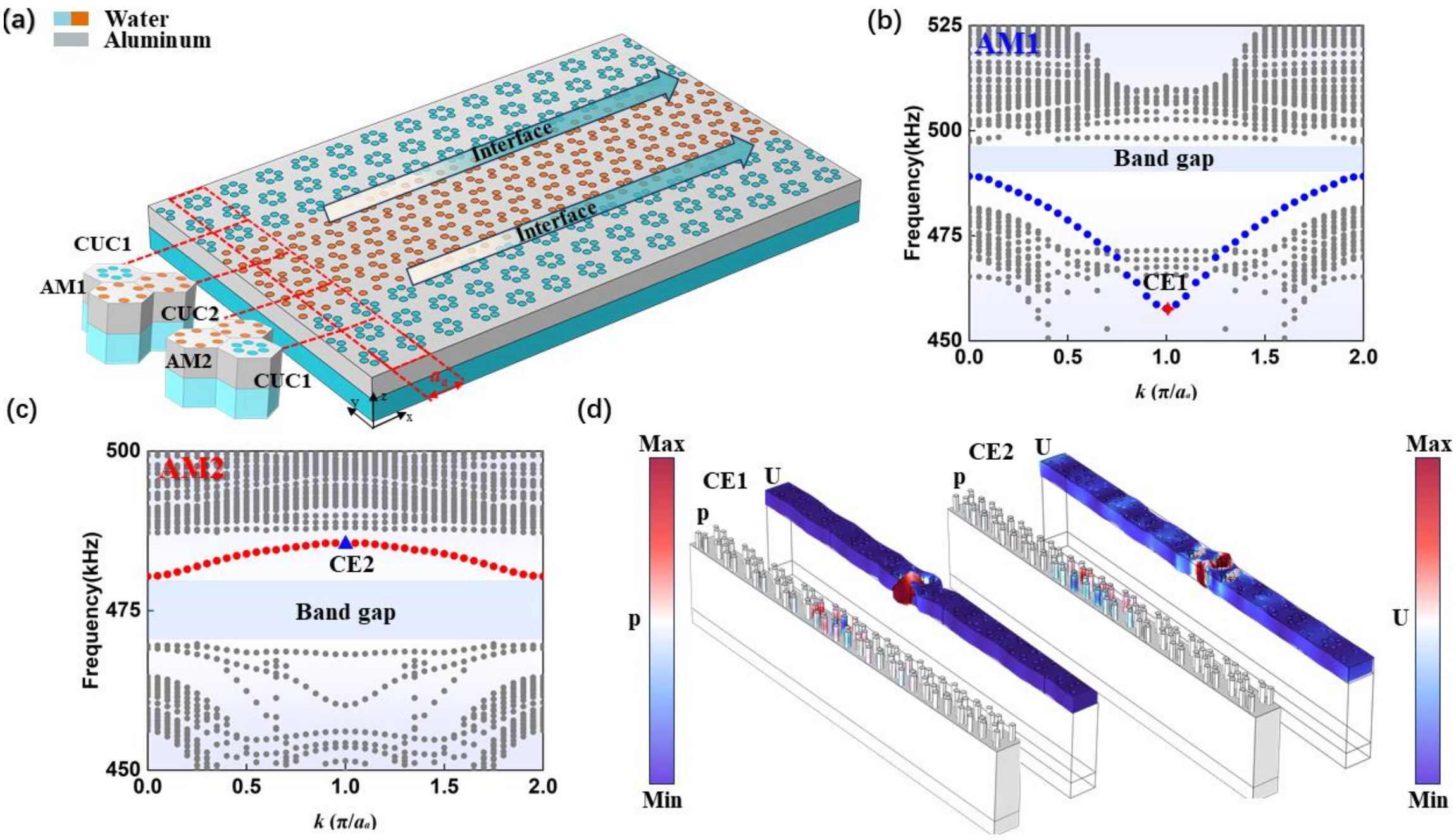

**Fig. 5**. QSHE ribbon supercells and dispersion relations. (a) Supercell composed of contracted CUC1 ($\Delta d = -0.26d$) and expanded CUC2 ($\Delta d = +0.15d$), forming the armchair interfaces AM1 and AM2. (b) The dispersion relations of the supercell containing the AM1 interface. The gray regions denote the bulk bands, the blue curve denotes the AM1 edge branch, and the blue shaded region marks the upper AM1 edge bandgap. (c) The dispersion relations of the supercell containing the AM2 interface. The red curve denotes the AM2 edge branch, and the blue shaded region marks the lower AM2 edge bandgap. CE1 and CE2 indicate the representative eigenmodes shown in (d). (d) Scalar acoustic pressure fields in water and vector elastic-displacement fields in the aluminum plate for CE1 and CE2. The simultaneous localization of the two fields at AM1 and AM2 demonstrates the coordinated confinement of the coupled acoustic and elastic responses by the QSHE edge states.

# 3. Topological Interface Transport and Higher-Order Corner States

To examine SSAW transport along the topological interfaces, two finite straight-channel models are constructed. The QVHE model in Fig. 6(a) consists of an upper UC1 domain ($\delta = -0.1$, blue) and a lower UC2 domain ($\delta = +0.1$, orange), which have opposite valley Chern indices and form a zigzag interface. The QSHE model in Fig. 6(c) combines the contracted CUC1 domain ($\Delta d = -0.26d$, blue) and the expanded CUC2 domain ($\Delta d = +0.15d$, orange), forming an armchair interface between the trivial and nontrivial pseudospin-Hall phases. In both models, a monopole source is positioned near the left end of the interface to excite the corresponding in-gap mode, and perfectly matched layers are applied at the exterior computational boundaries to absorb outgoing waves and suppress artificial reflections. Figure 6(b) shows the acoustic-energy distribution of the QVHE model at 482 kHz. The field launched by the source remains predominantly confined to the zigzag interface and propagates across the full channel, while its amplitude decays rapidly in the transverse direction with little penetration into the adjoining bulk domains. This response is consistent with the QVHE interface branch identified in Fig. 4 and confirms the excitation and transport of an interface-confined SSAW across the modeled channel. The acoustic energy distribution of the QSHE model at 501 kHz is presented in Fig. 6(d). The intensity maxima extend along the armchair interface and remain strongly confined between the two bulk domains. These characteristics stem from the topological protection provided by the difference in topological phases across the interface, ensuring robust acoustic signal transport along predefined paths.

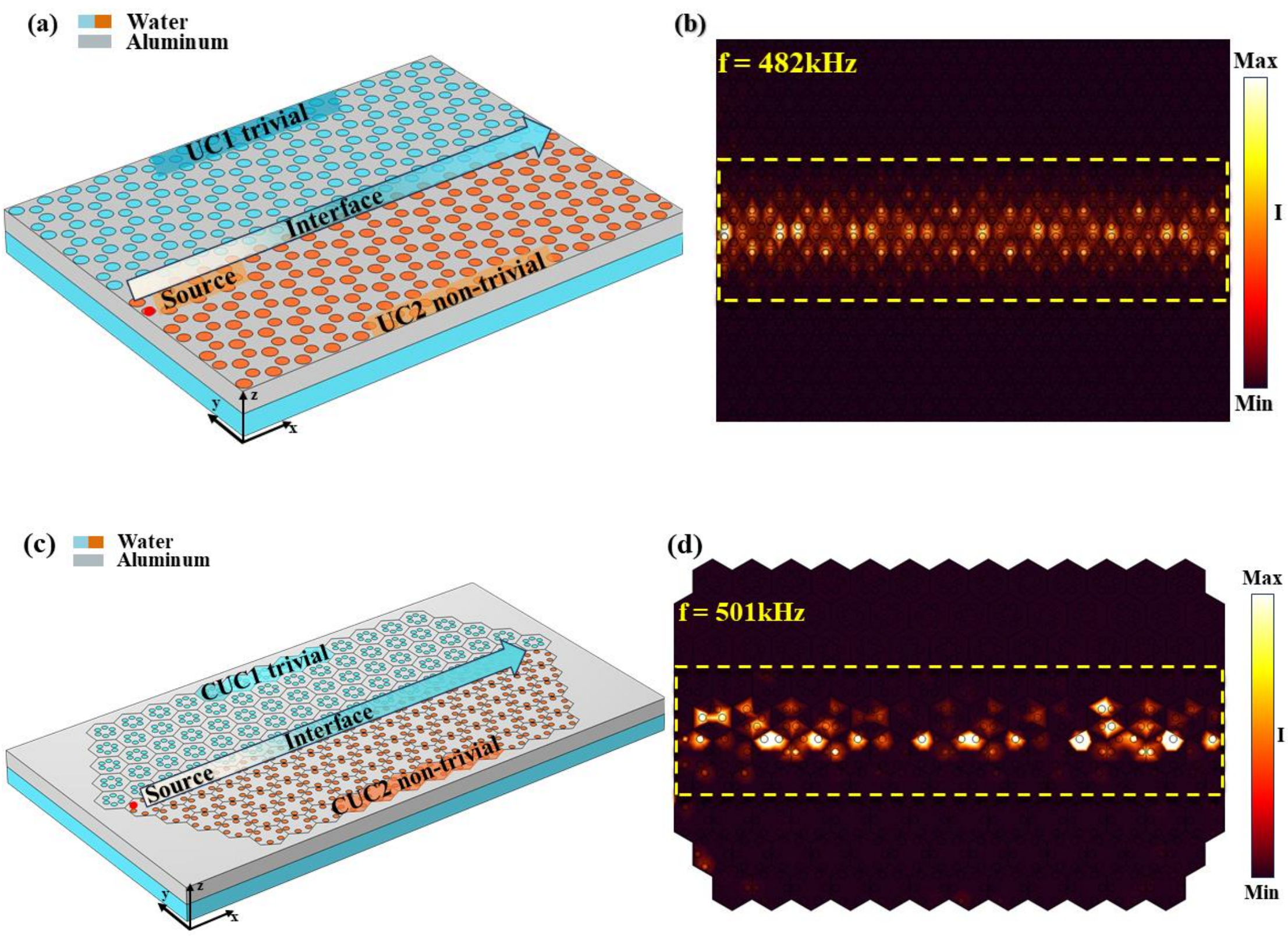

**Fig. 6.** Straight-interface transport of SSAWs in the QVHE and QSHE lattices. (a) Finite QVHE model formed by UC1 ($\delta = -0.1$) and UC2 ($\delta = +0.1$), which possess opposite valley Chern indices and form a zigzag interface. (b) Acoustic energy distribution at 482 kHz, showing propagation along the QVHE interface and transverse confinement within the dashed region. (c) Finite QSHE model formed by contracted CUC1 ($\Delta d = -0.26d$) and expanded CUC2 ($\Delta d = +0.15d$), which form an armchair interface between the two pseudospin-Hall phases. (d) Acoustic energy distribution at 501 kHz, showing the interface-confined QSHE edge mode.

To verify the higher-order corner states, a finite triangular metastructure is constructed by embedding a CUC2 domain (orange) within a CUC1 domain (blue), as shown in Fig. 7(a). CUC1 and CUC2 represent the trivial and nontrivial pseudospin-Hall phases, respectively. Three symmetry-related corners are formed in this metastructure. For $R = R_0$, with $R_0 = R_a = 0.26$ mm, the computed eigenfrequency spectrum in Fig. 7(b) separates the bulk, edge, and corner states. Three nearly degenerate corner eigenvalues appear near 522 kHz within the edge bandgap. Figure 7(c) shows the eigenfield of a representative corner mode at 522 kHz. The field maximum is located at one CUC1-CUC2 corner and decays rapidly both along the adjoining interfaces and into the surrounding bulk domains, confirming its zero-dimensional localization.

To assign distinct resonance frequencies to the four corner sites in the final device shown in Fig. 1, the hole radius R at each corner is used as a local tuning parameter. As shown in Fig. 7(d), increasing R/R0 from 1.0 to 1.3 monotonically reduces the corner-state frequency from approximately 522 to 504 kHz. This frequency-radius relation provides the design basis for frequency-selective corner localization. The corresponding eigenfields in Fig. S2 further show that the field remains concentrated at the designated corner over the investigated parameter range.

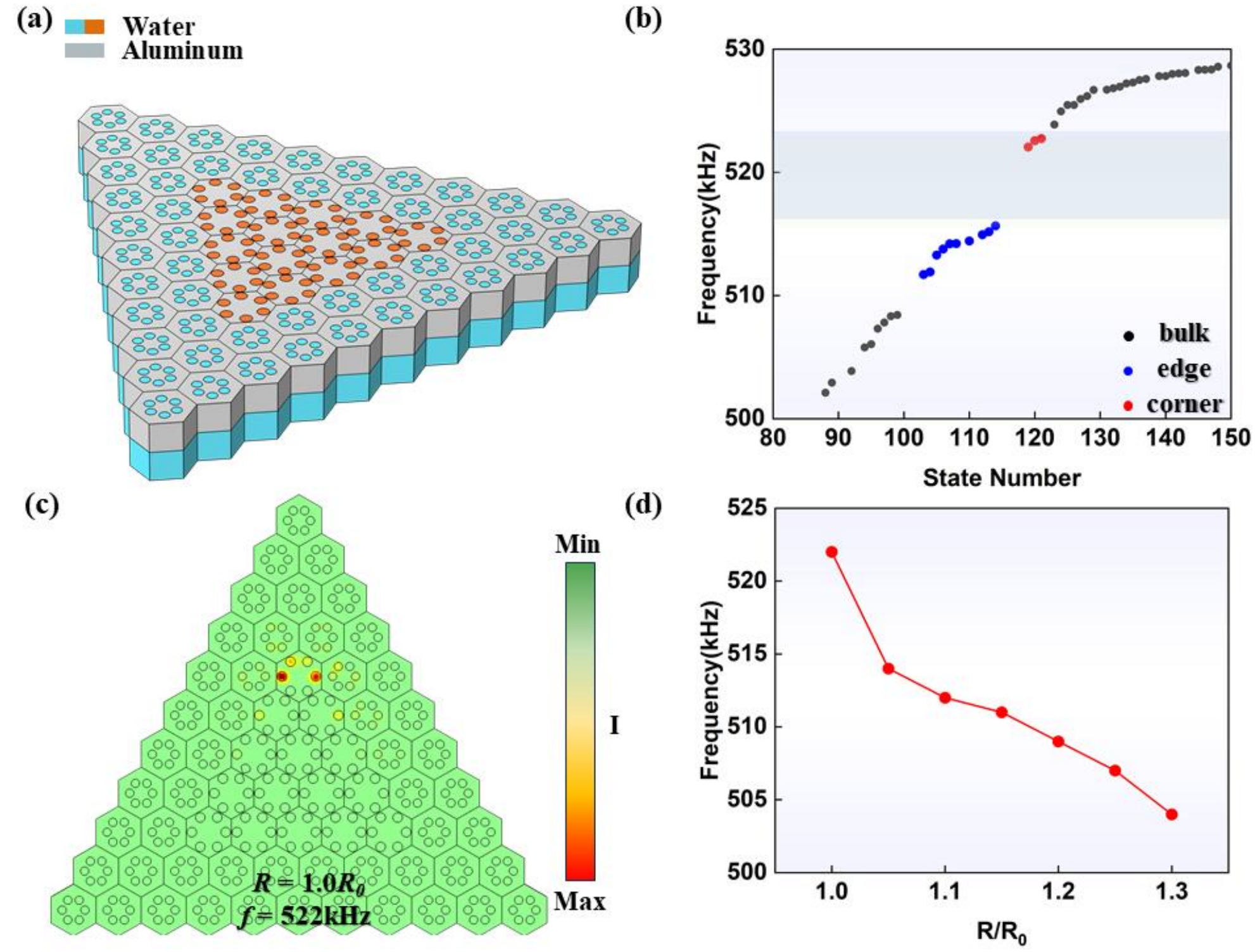

**Fig. 7.** Formation and geometric tuning of QSHE corner states. (a) Finite triangular metastructure formed by an inner CUC2 domain and an outer CUC1 domain, producing three symmetry-related corners along their interface. (b) Eigenfrequency spectrum for $R = R_0 = 0.26$ mm. Black, blue, and red markers denote bulk, edge, and corner states, respectively; the shaded region indicates the edge bandgap. (c) Acoustic-energy distribution of a representative corner state at 522 kHz for $R / R_0 = 1.0$, showing strong localization at one CUC1-CUC2 corner. (d) Corner-state frequency as a function of the hole radius $R / R_0$, demonstrating a monotonic frequency decrease with increasing R.

# 4. Topological rainbow demultiplexing and corner trapping

To realize QVHE-based spatial-frequency demultiplexing in an open underwater environment, a four-segment topological metadevice is constructed, as shown in Fig. 8(a). The segments, labeled A–D, are arranged sequentially along the propagation direction. In each segment, the upper UC1 domain and the lower UC2 domain possess opposite valley Chern indices and form a topological interface. By assigning different symmetry-breaking parameters $\delta$ ($\delta_{A1}$ = −0.1, $\delta_{B1}$ = −0.15, $\delta_{C1}$ = −0.2, and $\delta_{D1}$ = −0.25, $\delta_{A2}$ = 0.1, $\delta_{B2}$ = 0.15, $\delta_{C2}$ = 0.2, and $\delta_{D2}$ = 0.25) to these successive regions, the local interface-state dispersion and the associated band edge shift toward higher frequencies along the wave propagation direction, thereby establishing a piecewise spectral gradient along the propagation path.

Figure 8(b) illustrates the underwater acoustic energy distributions at excitation frequencies of 471, 474, 476, and 480 kHz, respectively. As the excitation frequency increases, the field-accumulation position shifts progressively downstream: the 471 kHz component is confined near the entrance region, whereas the 474, 476, and 480 kHz components reach successively more distant segments. This behavior is consistent with a progressive reduction in group velocity as each frequency component approaches the local interface-state band edge, resulting in frequency-dependent wave accumulation. The fields remain tightly confined in the transverse direction, confirming a direct frequency-to-position mapping and the spatial demultiplexing of the incident multifrequency signal along the QVHE interface.

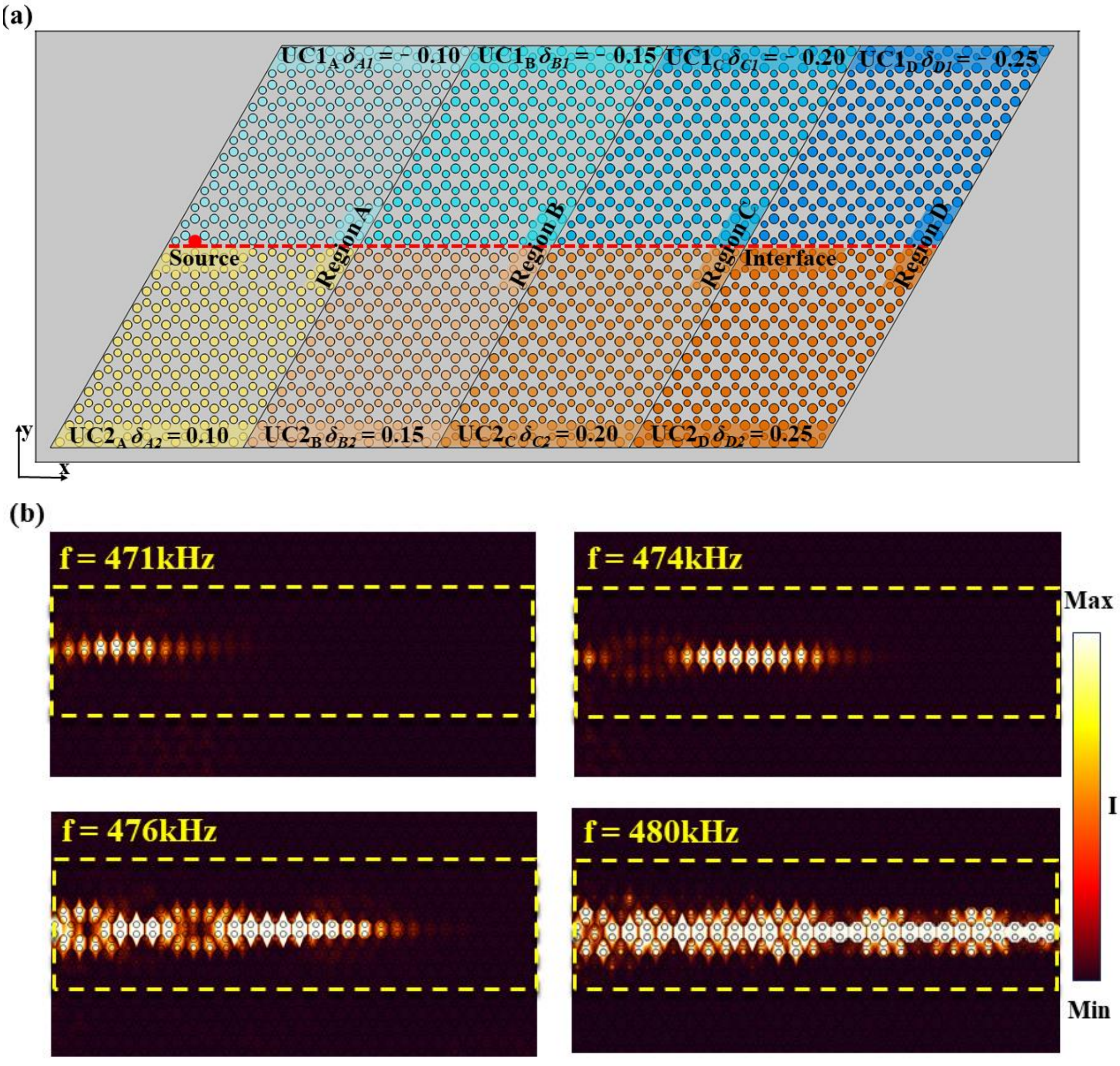

**Fig. 8.** QVHE-based topological rainbow demultiplexing. (a) Four-segment interface with stepwise perturbation

magnitudes $|\delta|$ = 0.10, 0.15, 0.20, and 0.25 in regions A-D, respectively. In each segment, UC1 and UC2 carry opposite valley Chern indices. (b) Acoustic energy distributions at 471, 474, 476, and 480 kHz, showing the downstream shift of the field-accumulation position with increasing frequency. The dashed rectangles indicate the interface channel.

The QSHE configuration extends the rainbow mechanism from one-dimensional edge localization to zero-dimensional corner confinement. As shown in Fig. 9(a), the open SSAW system consists of four domains, labeled I-IV. Domain I is formed by the trivial CUC1 phase, whereas domains II-IV are formed by the nontrivial CUC2 phase. Their arrangement creates a common edge-transport channel and four corner sites, C1-C4, at the lower vertices of domains II and III. These sites join boundaries with distinct edge terminations and edge bandgaps, providing the spectral and geometric conditions required for higher-order corner localization. The corner-hole radii are set to $R_{C1} = 1.2R_a$, $R_{C2} = 1.15R_a$, $R_{C3} = 1.10R_a$, $R_{C4} = 1.05R_a$, where $R_a = 0.26$ mm, to detune the four local corner resonances. The operating frequencies are selected within the propagating edge-state window so that the incident signals can first be delivered along the common interface.

Figure 9(b) presents the acoustic energy fields at 500.0, 501.4, 502.2, and 504.4 kHz. At all four frequencies, the wave launched from the same source initially follows the common edge channel. It is then transferred to and confined at C1, C2, C3, or C4 when the excitation frequency matches the corresponding corner resonance. The trapping position therefore shifts progressively away from the source from C1 to C4, demonstrating a transport-then-confinement process that combines edge-mediated delivery with frequency-selective remote corner localization. The weak field away from the selected corner indicates limited coupling to the other corner sites.

In conventional graded acoustic structures, abrupt variations in geometric parameters may act as defects or impedance mismatches, thereby inducing interfacial backscattering and reducing trapping efficiency and signal purity. However, in the proposed topological meta-device, the local resonant frequencies vary with the geometric parameters, while the topological interface states are maintained along the propagation path. This topological protection enables signals at different frequencies to traverse the graded waveguide with limited scattering before reaching their designated trapping sites. The resulting high-contrast spatial frequency separation provides a physical basis for broadband acoustic-energy harvesters and high-signal-to-noise-ratio sensor arrays in complex underwater environments.

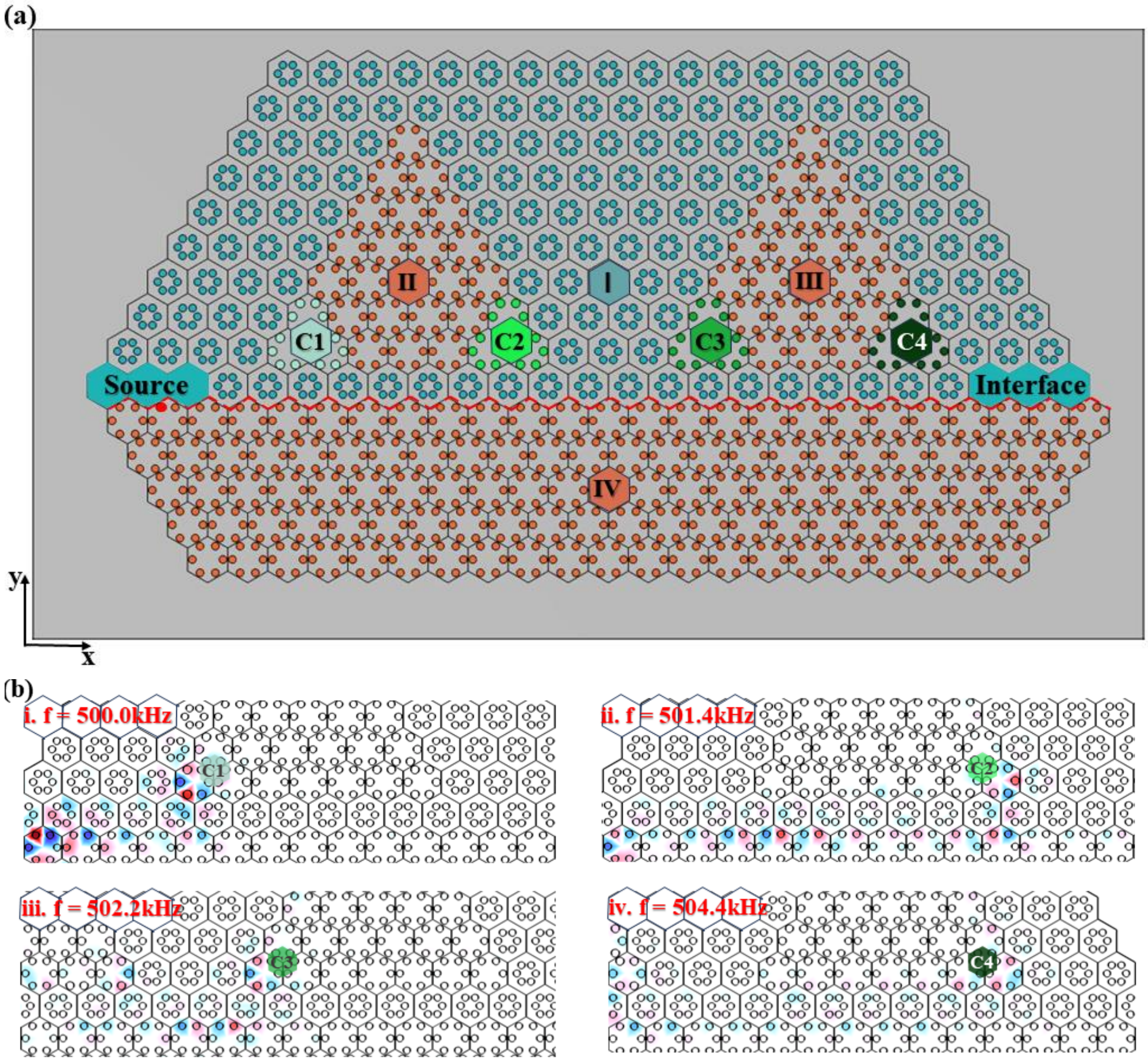

**Fig. 9.** QSHE-based edge-to-corner rainbow trapping. (a) Open SSAW system comprising four domains (I-IV), a common one-dimensional edge channel, and four locally tuned corner sites (C1-C4). The corner-hole radii are $R_{C1}$ = 1.20$R_a$, $R_{C2}$ = 1.15$R_a$, $R_{C3}$ = 1.10$R_a$, $R_{C4}$ = 1.05$R_a$ . (b) Acoustic energy distributions at 500.0, 501.4, 502.2, and 504.4 kHz, showing edge-mediated transport followed by frequency-selective localization at C1, C2, C3, and C4, respectively.

# 5. Defect tolerance of QVHE- and QSHE-based rainbow trapping

To evaluate the tolerance of the proposed topological rainbow-trapping mechanisms to structural imperfections, defect-containing models are constructed for both the QVHE- and QSHE-based systems. The selected defects include local variations in hole size and missing holes along the propagation channels or near the trapping sites. The corresponding pressure-field distributions are examined to determine whether frequency-dependent transport and localization are preserved.

As shown in Fig. 10(a), three defects are introduced along the graded interface of the QVHE rainbow device. Defect A enlarges one boundary hole in the second segment, defect B removes an interface hole in the third segment, and defect C enlarges one boundary hole in the fourth segment.

These perturbations locally modify the interface geometry without changing the valley-Hall phases of the adjoining bulk domains. The acoustic energy distributions in Fig. 10(b) show that the 474 kHz component remains confined to its expected upstream accumulation region, whereas the 480 kHz component propagates through the perturbed interfaces and reaches the downstream segment. In both cases, the field remains concentrated near the domain wall with limited radiation into the bulk. The preservation of the frequency-dependent accumulation pattern provides qualitative evidence that the QVHE rainbow response is tolerant to the examined defects, provided that intervalley scattering is not strongly induced.

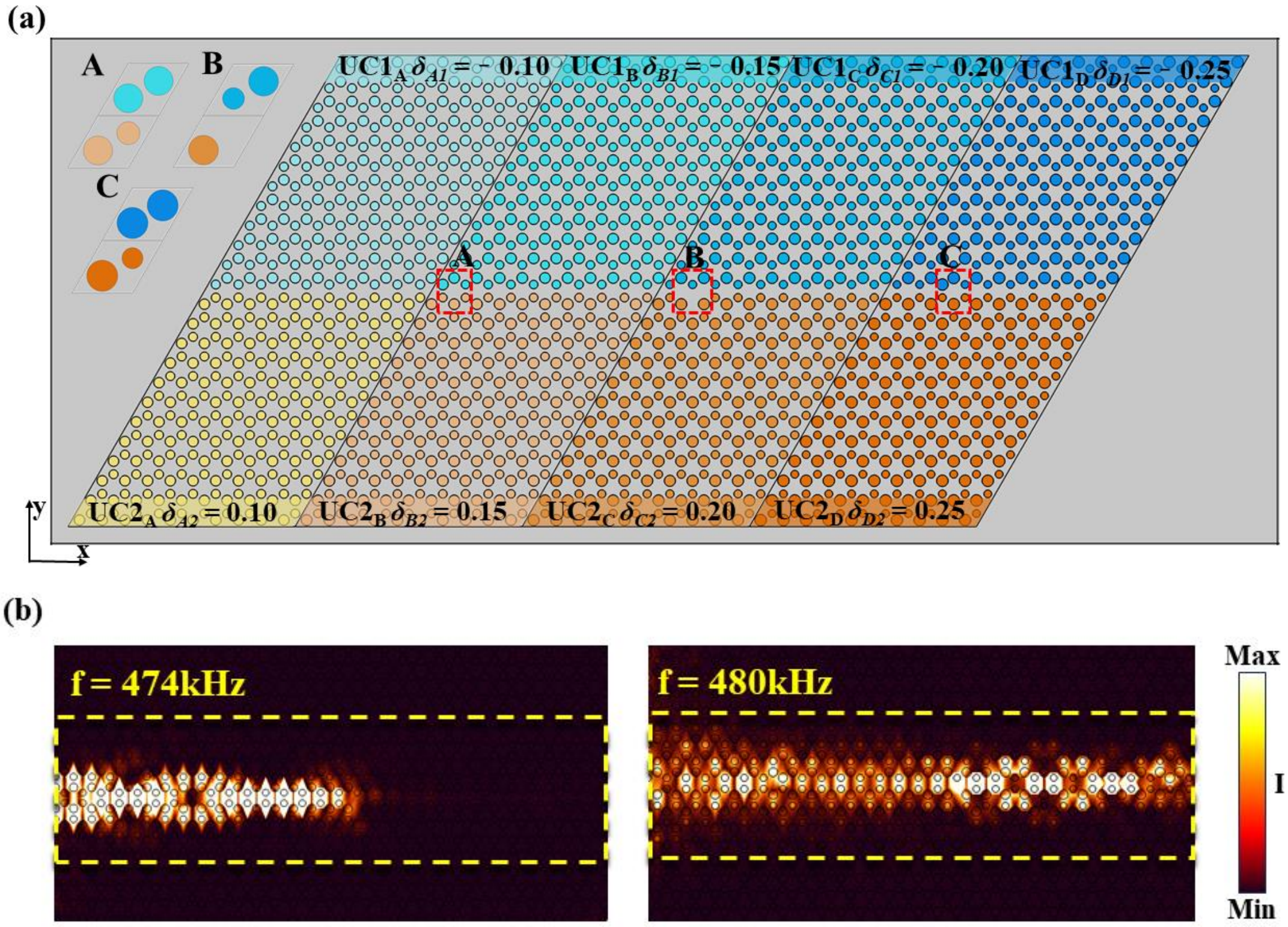

**Fig. 10.** Defect-tolerance analysis of the QVHE-based topological rainbow device. (a) Graded interface containing three local perturbations: enlarged one boundary hole in the second segment (A), an interface-hole vacancy in the third segment (B), and enlarged one boundary hole in the fourth segment (C). The red dashed boxes mark the defect locations. (b) Acoustic energy distributions at 474 and 480 kHz, demonstrating preserved interface confinement and frequency-dependent field accumulation in the presence of the defects.

The defect tolerance of the QSHE edge-to-corner mechanism is examined using the perturbed structure in Fig. 11(a). Defect A introduces a hole vacancy in the second corner site C2, defect B contracts selected holes in the fourth corner site C4, and defect C removes holes from the common edge-transport channel. Despite these local perturbations, the pressure fields in Fig. 11(b) retain the intended transport-then-confinement behavior. At 501.4 kHz, the guided field is delivered to and localized at C2, while at 504.4 kHz it propagates farther along the same edge channel and is concentrated at C4. The weak response at the nonresonant corner sites indicates that the frequency-selective trapping sequence is preserved. These results provide qualitative evidence that both edge-mediated delivery and remote corner localization remain functional under the tested defects, as long as the perturbations do not close the relevant bulk and edge gaps or strongly mix the pseudospin channels.

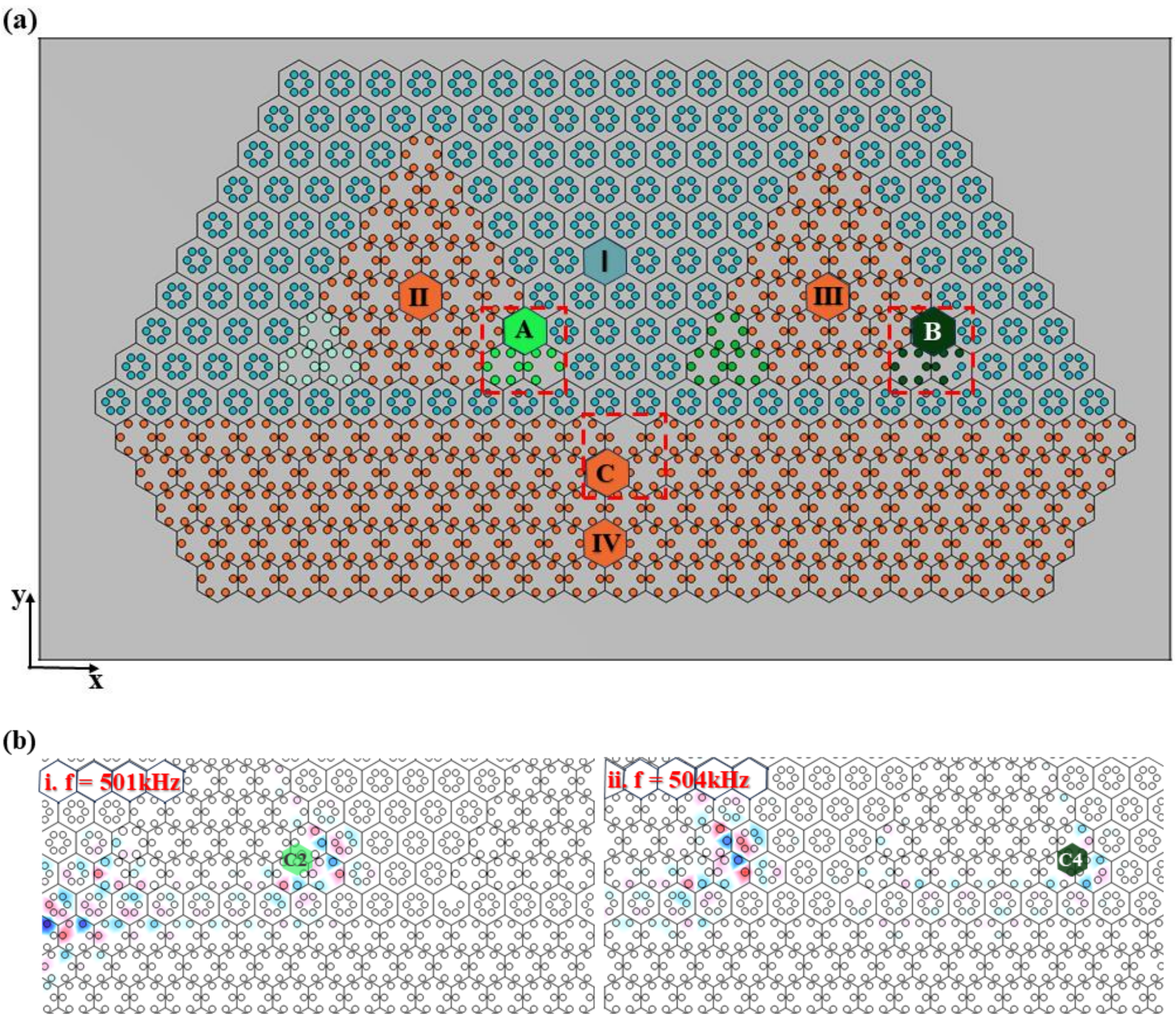

**Fig. 11.** Defect-tolerance analysis of the QSHE edge-to-corner trapping device. (a) Finite structure containing a hole vacancy in C2 (A), locally contracted holes in C4 (B), and hole vacancies along the common edge channel (C). The red dashed boxes mark the perturbed regions. (b) Acoustic pressure fields at 501.4 and 504.4 kHz, showing edge-mediated transport followed by frequency-selective localization at C2 and C4, respectively.

# 6. Experimental validation

To experimentally verify underwater topological rainbow separation and trapping, phononic-crystal aluminum-plate samples I and II were fabricated by precision computer numerical control (CNC) machining. Sample I implements the QVHE graded interface, whereas sample II implements the QSHE edge-to-corner configuration. The experimental configurations are shown in Figs. 12(a, b) and Figs. S3(a, b). Each sample was securely clamped at the bottom to ensure mechanical stability. To reproduce the underwater operating environment and approximate 2D propagation, the samples were immersed in an anechoic water tank. Underwater absorbing wedges were arranged around the samples to suppress reflections from the tank boundaries and approximate open radiation conditions.

A piezoelectric actuator (AE0203D04DF, 5 mm in length) was bonded near the left entrance of the interface channel and driven by a Tektronix AFG31000 signal generator through an Aigtek ATA-2022H power amplifier. The near-surface acoustic pressure was measured using a miniature needle hydrophone mounted on a motorized three-dimensional translation stage. The hydrophone was scanned point by point over a predefined plane close to the structured surface. A computer-based control program coordinated the excitation, stage motion, and signal acquisition through an NI PXIe-5160 digitizer.

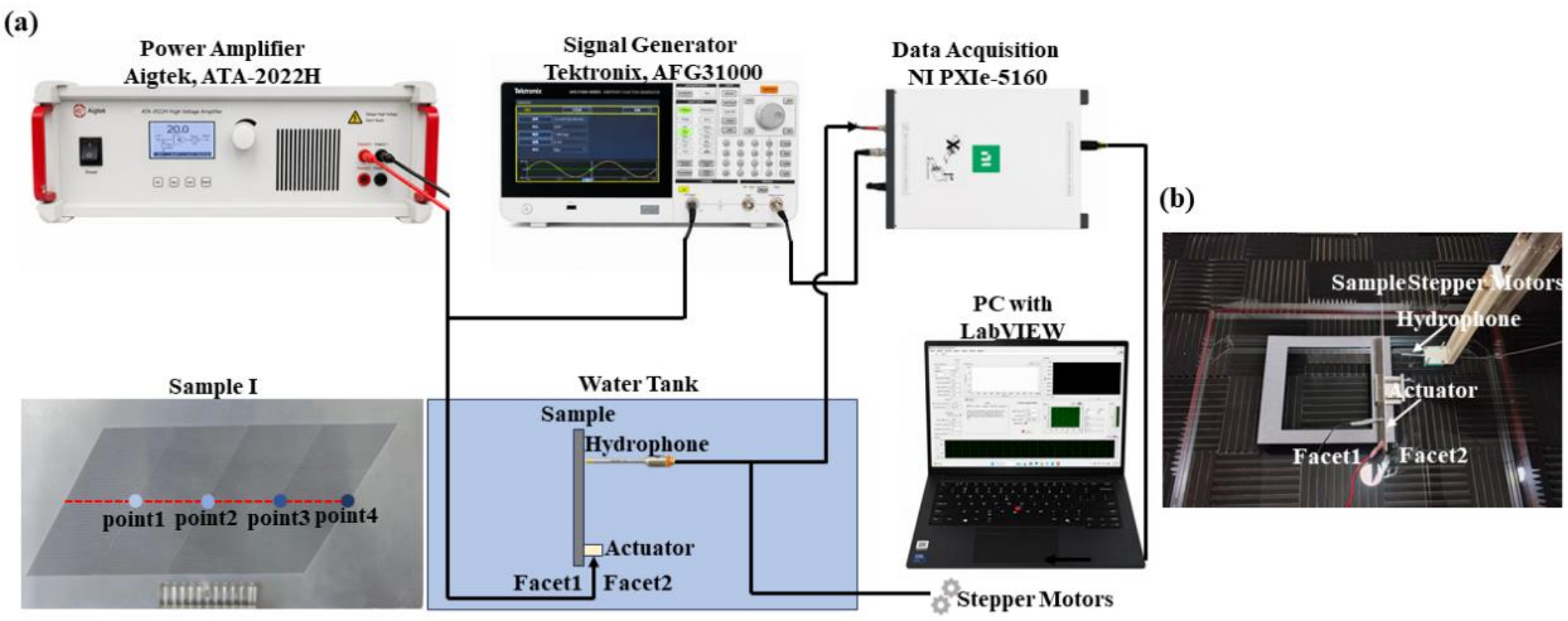

**Fig. 12.** Experimental platform for underwater SSAW measurements. (a) Instrumentation and signal-flow diagram, including the waveform generator, power amplifier, piezoelectric actuator, water tank, scanning hydrophone, NI PXIe-5160 digitizer, motorized stage, and computer-based control system. The inset marks the four observation points on sample I. (b) Photograph of the experimental arrangement, showing the submerged sample, actuator, scanning hydrophone, supporting facets, and stepper-motor positioning system.

The experiments evaluate the transmission and localization characteristics in terms of both frequency-domain responses and spatial-field distributions. The local spectral responses were first measured at four positions along each sample. The excitation frequency was swept across the target band, and the pressure amplitude was recorded at each observation point. For sample I, the spectra in Fig. 13(a) show position-dependent transmission windows. From point 1 to point 4, the lower-frequency boundary of the enhanced-response region shifts progressively upward, with approximate response intervals of 470-482, 472-482, 476-482, and 478-482 kHz, respectively. This systematic shift is consistent with the spatial variation of the local QVHE interface-state dispersion. For sample II, the spectra measured at C1-C4 in Fig. 13(c) exhibit four distinct resonances centered at approximately 500.0, 501.4, 502.2, and 504.4 kHz, confirming that the local corner geometries produce spectrally separated corner responses. The measured ordering agrees with the numerical prediction; residual frequency shifts can arise from machining tolerances, clamping conditions, water-property variations, and measurement uncertainty.

Two-dimensional scans were subsequently performed under single-frequency harmonic excitation to visualize the frequency-dependent localization. For sample I, the measured acoustic pressure-amplitude maps at 472, 474, 476, and 480 kHz are presented in Fig. 13(b). The field remains confined to the QVHE interface, while the longitudinal accumulation position moves progressively away from the source as the frequency increases. The weak response beyond each accumulation region and the limited transverse leakage experimentally demonstrate spatial-frequency demultiplexing along the graded interface. For sample II, the fields in Fig. 13(d) first follow the common QSHE edge channel and are then concentrated at C1, C2, C3, and C4 at the corresponding excitation frequencies. These measurements reproduce the transport-then-confinement process and verify frequency-selective remote corner localization. Together, the results validate QVHE-based rainbow demultiplexing and QSHE-based edge-to-corner trapping in the open underwater SSAW platform, supporting its potential for frequency-selective sensing and acoustic-energy concentration.

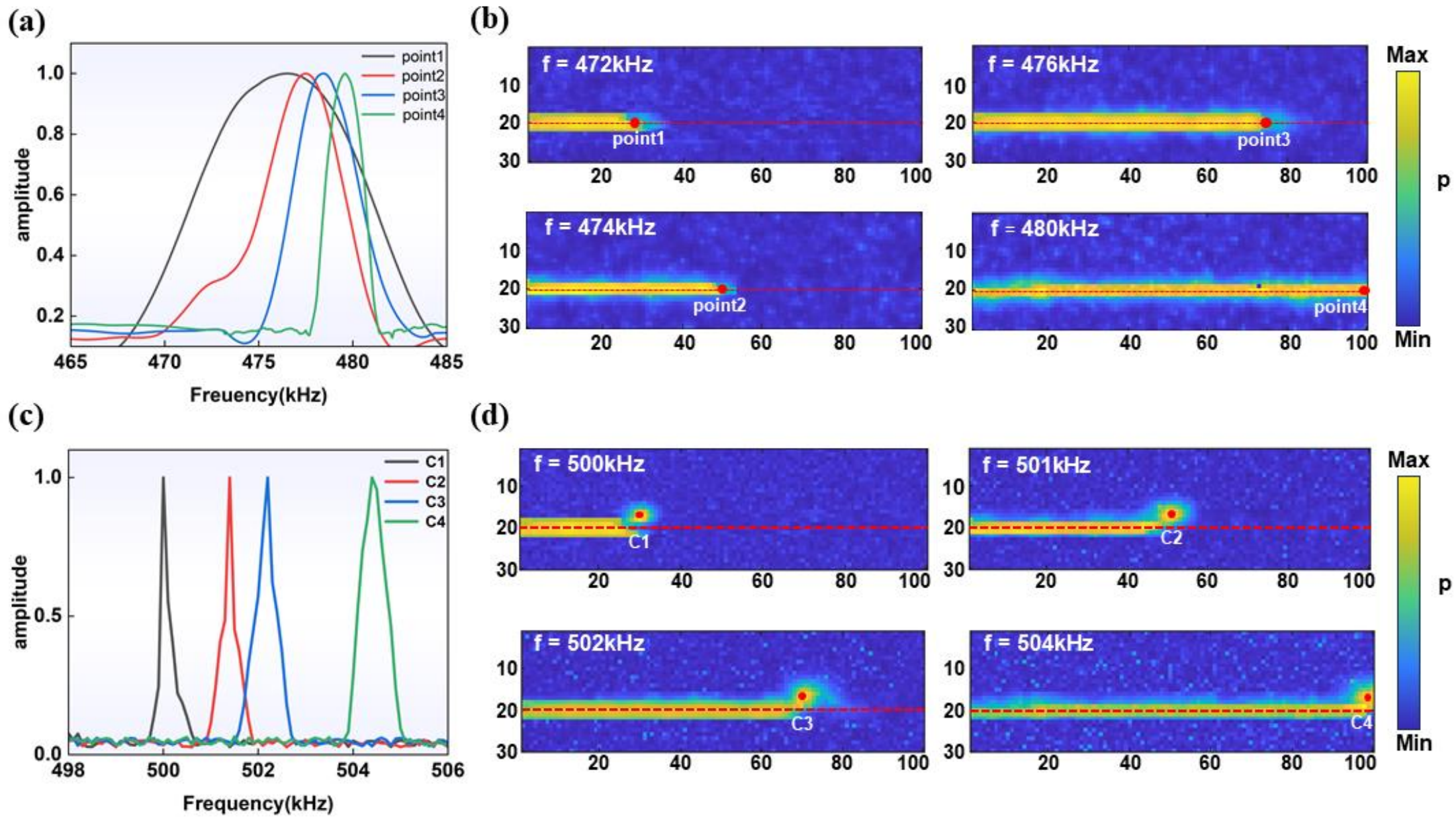

**Fig. 13.** Experimental validation of QVHE-based rainbow demultiplexing and QSHE-based edge-to-corner trapping. (a) Pressure amplitude spectra measured at points 1–4 along the graded QVHE interface of sample I. (b) Measured acoustic pressure-amplitude maps at 472, 474, 476, and 480 kHz, showing the frequency-dependent downstream shift of the field-accumulation position. (c) Pressure amplitude spectra measured at corner sites C1–C4 of sample II. (d) Measured acoustic pressure-amplitude maps at the indicated excitation frequencies, showing transport along the common QSHE edge channel followed by frequency-selective localization at C1–C4. The red dashed lines in (b) and (d) denote the corresponding edge-transport channels.

# 7. Conclusion

In conclusion, we established two complementary underwater topological SSAW platforms based on the acoustic analogues of the QVHE and QSHE. Calculations of the fully coupled fluid–solid dispersion relations established the bulk topological properties of the two systems. In the QVHE lattice, breaking inversion symmetry lifts the Dirac degeneracy at the **K** valley and produces two domains with opposite valley Chern indices. In the QSHE lattice, the *p*–*d* band inversion at the **Γ** point distinguishes the contracted and expanded pseudospin-Hall phases. Ribbon-supercell calculations subsequently identified gap-spanning interface states in the QVHE system and termination-dependent edge states in the QSHE system. The corresponding modal fields demonstrate that the scalar acoustic pressure response in water and the vector elastic-displacement response in the aluminum plate are simultaneously confined near the interfaces, confirming the coordinated localization of the coupled fluid–solid modes. Finite-structure calculations further verified interface-guided propagation, higher-order corner localization, and geometric tuning of the corner-state eigenfrequencies, thereby establishing the spectral basis for edge-to-corner coupling.

These topological states were subsequently used to implement two distinct rainbow mechanisms. For the QVHE device, a stepwise spatial variation of the symmetry-breaking parameter shifts the local interface-state dispersion, mapping different frequency components to different accumulation positions along the same propagation path. For the QSHE device, multifrequency signals are first transported through a common edge channel and are then selectively coupled into remote corner states whose resonances are set by the local geometry. This transport-then-confinement process separates the delivery and capture functions while retaining frequency selectivity. Tests containing structural perturbations showed that the guided fields remained confined to the designed interfaces around the examined defects. Measurements on the fabricated QVHE and QSHE samples reproduced the predicted position-dependent spectral responses and spatial pressure distributions, experimentally confirming both interface-based rainbow demultiplexing and frequency-selective corner trapping in an open underwater configuration.

Beyond the present demonstrations, the integration of passive spectral routing, topological transport, and subwavelength localization provides a basis for compact underwater spectral analyzers, distributed frequency-selective sensor arrays, multiplexed acoustic transducers, and wave-based analog preprocessing. Physical-layer separation of characteristic frequencies may facilitate underwater target recognition while reducing downstream processing requirements. Frequency-selective corner localization could also concentrate ambient acoustic energy at integrated transducers for energy harvesting. The same frequency-to-position mapping may support parallel spectral monitoring and channel-selective underwater communication. Future works will focus on dynamic tunability, broader and lower-frequency operation, loss reduction, conversion efficiency, and validation under realistic marine conditions.

## CRediT authorship contribution statement

**Cheng Lin:** Conceptualization, Methodology, Software, Investigation, Writing – original draft. **Yangkai Liu:** Methodology, Validation, Supervision. **Tuo Liu:** Software, Investigation, Validation. **Yi Zhang:** Validation, Resources, Writing – review & editing. **Haiyan Fan:** Supervision, Project administration Funding acquisition, Writing – review & editing. **Hui Zhang:** Conceptualization, Supervision, Project administration, Funding acquisition, Writing – review & editing.

## Declaration of competing interest

The authors declare that they have no known competing financial interests or personal relationships that could have appeared to influence the work reported in this paper.

## Acknowledgments

This work was supported by the National Natural Science Foundation of China (Grant Nos. 52501418 and 12574493), the Natural Science Foundation of Jiangsu Province (Grant No. BK20251266), and the Advanced Ocean Institute of Southeast University, Nantong (MP202601).

# 【参考文献】